\documentclass[%
reprint,
superscriptaddress,
showkeys,
amsmath,amssymb,
aps,
twocolumns]{revtex4-2}

\usepackage{amssymb}
\usepackage{amsmath}
\usepackage{amsthm}
\usepackage{natbib}
\usepackage{eucal}
\usepackage{mathrsfs}
\usepackage{graphicx}
\usepackage{dcolumn}
\usepackage{color}
\usepackage{bm}
\usepackage{hyperref}
\usepackage[usenames,dvipsnames,x11names,table]{xcolor}
\usepackage{balance}
\usepackage{dsfont}
\usepackage{tikz}
\usetikzlibrary{shapes}
\usepackage{multirow}
\usepackage{tabularx}
\usepackage{enumitem}
\usepackage{cancel,soul}

\setlist[itemize]{align=parleft,left=0pt..1em}
\newcommand{\Ra}{\operatorname{Ra}}
\newcommand{\Nu}{\operatorname{Nu}}

\makeatletter
\let\@internalcite\cite
\def\cite{\def\citeauthoryear##1##2{##1, ##2}\@internalcite}
\def\shortcite{\def\citeauthoryear##1##2{##2}\@internalcite}
\def\@biblabel#1{\def\citeauthoryear##1##2{##1, ##2}[#1]\hfill}
\makeatother
\begin{document}
\preprint{APS/123-QED}
\title{Rayleigh-B\'enard thermal convection in emulsions: a short review}
\author{Francesca Pelusi}
\email{francesca.pelusi@cnr.it}
\affiliation{Istituto per le Applicazioni del Calcolo, CNR - Via Pietro Castellino 111, 80131 Naples, Italy}
\author{Andrea Scagliarini}
\affiliation{Istituto per le Applicazioni del Calcolo, CNR - Via dei Taurini 19, 00185 Rome, Italy}
\affiliation{INFN, Sezione Roma ``Tor Vergata", Via della Ricerca Scientifica 1, 00133 Rome, Italy}
\author{Mauro Sbragaglia}
\affiliation{Department of Physics \& INFN, Tor Vergata University of Rome, Via della Ricerca Scientifica 1, 00133 Rome, Italy}
\author{Massimo Bernaschi}
\affiliation{Istituto per le Applicazioni del Calcolo, CNR - Via dei Taurini 19, 00185 Rome, Italy}
\author{Roberto Benzi}
\affiliation{Sino-Europe Complex Science Center, School of Mathematics \\North University
of China, Shanxi, Taiyuan 030051, China}
\affiliation{Department of Physics \& INFN, Tor Vergata University of Rome, Via della Ricerca Scientifica 1, 00133 Rome, Italy}
\date{\today}
\begin{abstract}
\noindent Thermally driven emulsions arise in a broad range of natural and industrial contexts, yet their fundamental physical understanding remains only partially established. Emulsions exhibit a complex, concentration-dependent rheology, ranging from Newtonian (dilute emulsions) to yield-stress (concentrated emulsions). In buoyancy-driven flows, the complex structure and rheology of the emulsion are strongly coupled to convective flows, giving rise to fascinating and non-trivial phenomena involving stability, transient dynamics, and morphological evolution of the system. We review recent progress on thermally driven emulsions in the celebrated Rayleigh–B\'enard configuration, offering new perspectives on the behaviour of soft materials in thermal convection.

\end{abstract}


\maketitle

\section{Introduction}
\begin{figure*}[th!]
\centering
\includegraphics[width=.95\linewidth]{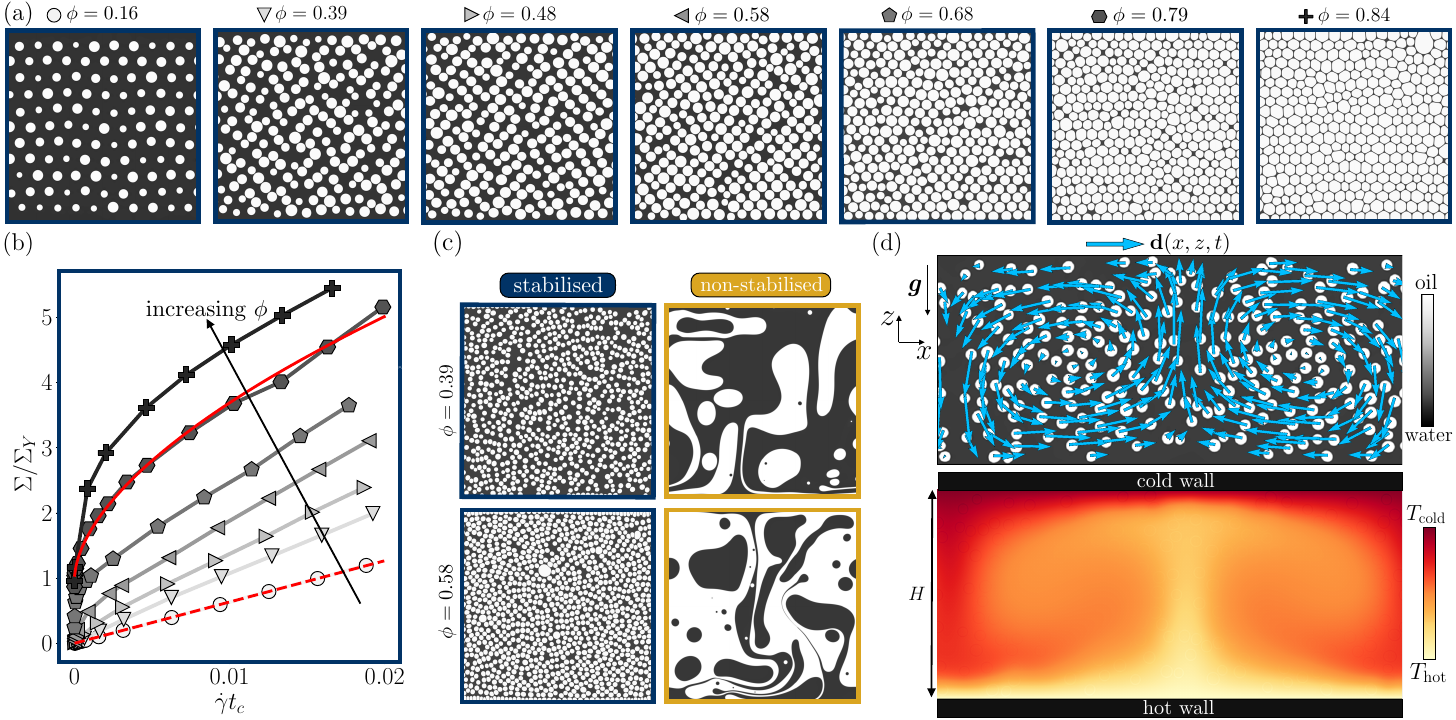}
\caption{Panel (a): Snapshots of oil-in-water emulsions with different values of the volume fraction of the initially dispersed phase, $\phi$. 
Panel (b): Rheological flow-curve relating the shear stress $\Sigma$ (in units of $\Sigma_Y$, the yield stress at $\phi=0.79$) to the shear rate $\dot{\gamma}$ [in units of $t_c^{-1}=\Gamma/(\eta R)$, where $\Gamma$ is the surface tension, $\eta$ the dynamic viscosity and $R$ the mean droplet radius], for the emulsions shown in panel (a); the dashed red line represents the Newtonian relation $\Sigma = \eta_{\text{eff}}(\phi)\dot{\gamma}$, with $\eta_{\text{eff}}(\phi) = \eta\left(1+\frac{7}{4}\phi\right)$ (with $\phi=0.16$), whereas the solid red line refers to the Herschel-Bulkley fit (see text). Panel (c): Snapshots of stabilised (i.e., proper emulsions, left column) and non-stabilised (right column) liquid-liquid dispersions at $\phi<0.5$ (top row) and $\phi>0.5$ (bottom row). Panel (d): oil-in-water emulsion in a 2D RB cell, confined between a lower hot and an upper cold wall and under the effect of buoyancy forces. Light blue arrows highlight droplet displacements, $\mathrm{\bf d}  (x,z,t)$, at a given time during convective dynamics, marking the convective rolls. The corresponding thermal plume is also shown.}
\label{fig.1}
\end{figure*}

Emulsions are mixtures of two immiscible liquids that are encountered in a variety of contexts, ranging from industry~\cite{Khan11,Goodarzi19} and technology~\cite{MartiMestres02,Tadros09} to fundamental research~\cite{Balmforthetal14,Bonn17}. At mesoscale, emulsions appear as a collection of microscopic droplets of a dispersed phase (say oil) in another continuous phase (say water); the presence of surfactants at the droplets interfaces is responsible for the emergence of a positive disjoining pressure~\cite{Tcholakova2004,Ravera2021} that inhibits droplets coalescence and, in turn, results in non trivial collective behaviours with a complex rheology depending on the volume fraction of the dispersed phase: dilute emulsions behave as Newtonian fluids with augmented viscosity~\cite{Taylor32,Zinchenko84}, whereas, as droplet concentration 
increases, non-Newtonian behaviours emerge, ranging from shear thinning~\cite{Pal2000}, to yield-stress rheology~\cite{Balmforthetal14,Bonn17}, in jammed emulsions [see Fig.~\ref{fig.1}(a)-(c)].\\ 
Studies of emulsions fluid dynamics are typically performed under isothermal conditions and restricted to their characterization under rheometric~\cite{Barnes94,Derkach09,Balmforthetal14,Bonn17}, homogeneous shear-driven~\cite{Loewenberg96,Yu2002,Zinchenko02,Benzietal14}, or pressure-driven flows~\cite{Dollet15,Pelusi24rheology}, also in turbulent conditions~\cite{Biferale11,Yi21,Yi24}. 
Nevertheless, thermal flows of emulsions are relevant in geophysical contexts, where they are employed as analogue systems to model buoyancy-driven processes such as lava flows~\cite{Griffiths2000,Lavallee15} and Earth's mantle convection~\cite{Montelli06,French15,Davaille2018}. In these settings, the interplay between temperature gradients, complex rheology and morphology evolution crucially affects the resulting heat transport and flow organisation. 
Despite this relevance, the behaviour of emulsions under thermal forcing remains only partially explored. This gap is mainly due to the intrinsic multiscale nature of the problem, which involves the coupling between interfacial dynamics and large-scale buoyancy-driven motions~\cite{Karimfazli16,Metivier17}.\\ 
When buoyancy forcing exceeds a critical threshold, a fluid layer confined between a bottom hot plate and a top cold plate [see Fig.~\ref{fig.1}(d)] is prone to the well-known Rayleigh-Bénard (RB) instability and transition to convection~\cite{Benard1900,Rayleigh1916}. For larger forcing, the system becomes turbulent, displaying multiscale dynamics and a complex interplay between mean wind and boundary layer physics (including fluid-wall interactions)~\cite{Grossmann99,Ahlers09,Lohse10,Chilla12,Stevens13}. This phenomenology is further enriched by the presence of another dispersed phase that couples and interacts with the thermal flow, as witnessed by many studies in the literature~\cite{Orestaetal09,Lakkaraju13,Biferale12,Garoosi21,Santos21,Liu21a,Liu22,Brandt24,Mangani24}, which, however, considered only dilute situations that display a Newtonian rheology. At high volume fractions, emulsions indeed exhibit non-Newtonian rheology, further complicating the problem. Some theoretical and numerical works approach the problem at the level of hydrodynamic equations with prescribed constitutive laws for the stress tensor~\cite{Zhang06,BalmforthRust09,Vikhansky09,Vikhansky10,AlbaalbakiKhayat11,Turanetal12,Massmeyer13,Hassanetal15,Karimfazli16}. However, especially at higher Reynolds numbers and moderate-to-large volume fractions, the dynamics is strongly affected by the balance of flow-induced breakup and coalescence of droplets, which, in turn, dictates the local and global rheological properties of the material~\cite{PelusiPRE25}. Moreover, neglecting the finite size of the droplets has a non-trivial impact on the cooperativity mechanisms generated when configurations of microscopic constituents rearrange~\cite{Goyon08,Goyon10}, influencing the dynamics of the thermal flow~\cite{Davailleetal13}. In summary, a theoretical and computational framework able to cope with resolved finite-sized droplets and interface dynamics, featuring complex rheology and break-up/coalescence dynamics, is, therefore, in order. With this aim, novel mesoscale methods have recently been developed~\cite{Sbragagliaetal12,Dollet15,Benzietal14}  through which recent works~\cite{PelusiSM21,PelusiSM23,PelusiPRL24,PelusiPRE25,PelusiPRF25} provided a detailed characterization of the dynamics of emulsions with finite-sized droplets, discussing how the non-trivial interplay between the emulsion rheology and the buoyancy forces gives rise to an unprecedented richness of dynamical regimes. The present contribution critically summarises these findings.
\section{Methods}
\begin{figure}
\centering
\includegraphics[width=.95\linewidth]{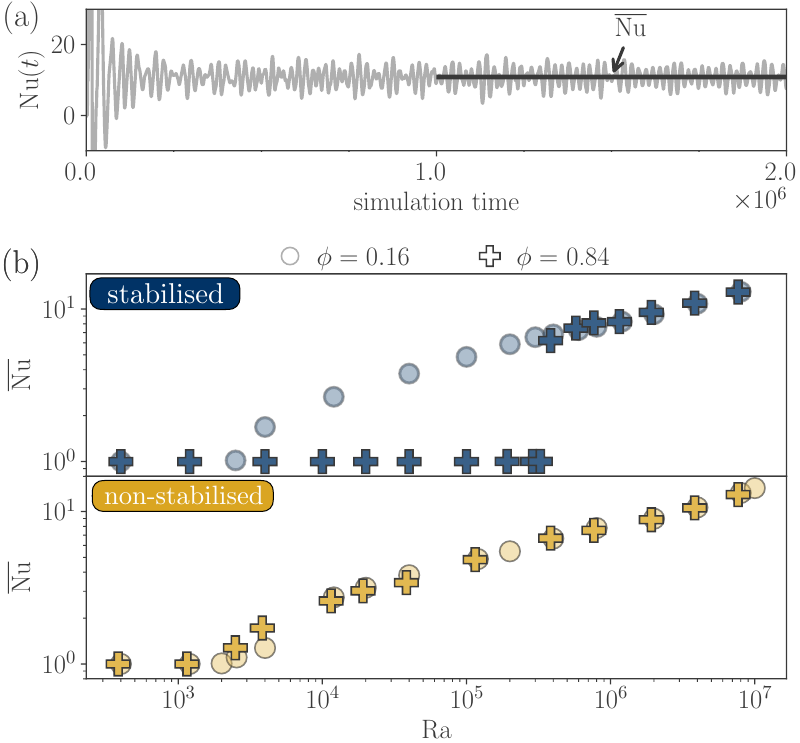}
\caption{Analysis of heat transfer at macroscopic scales. Panel (a): time evolution of the Nusselt number $\Nu(t)$, highlighting the time average of the Nusselt number $\overline{\Nu}$ over the statistically steady state [see Eq.~\eqref{eq:Nu}]. Data refer to an emulsion with  $\phi=0.16$ for a value of the Rayleigh number $\Ra \approx 4 \times 10^6$. Time is reported in simulation units. Panel (b): $\overline{\Nu}$, as a function of $\Ra$ [see Eq.~\eqref{eq:Ra}] for $\phi = 0.16$ and $\phi = 0.84$. We report cases of stabilised (i.e., proper emulsions) and non-stabilised liquid-liquid dispersions.}
\label{fig.2}
\end{figure}

All numerical studies ~\cite{PelusiSM21,PelusiSM23,PelusiPRL24,PelusiPRE25,PelusiPRF25} on thermally driven emulsions discussed in this review are based on TLBfind~\cite{TLBfind22}, an open-source code which implements state-of-the-art 2D multicomponent lattice Boltzmann methods~\cite{Benzi92,Succi18}. Within this framework, each of the two fluid components (``oil" and ``water") is represented by its own kinetic distribution function, whose evolution gives rise to phase-separating diffuse interfaces via intra-component interactions~\cite{ShanChen93,ShanChen94}. Additional interactions can be incorporated to generate a positive disjoining pressure~\cite{Sbragagliaetal12}, thereby mimicking the action of surfactants and enabling the stabilisation of emulsions against full phase separation. The reader is referred to dedicated works~\cite{Sbragagliaetal12,Dollet15,TLBfind22} for detailed technical characterization of the models. The employed methodology is particularly well suited for simulating 2D emulsions under RB thermal convection, i.e., when the system is subjected to a fixed temperature difference $\Delta T$ between horizontal plates, of length $L$, at a distance $H$ [heated from below and cooled from above, see Fig.~\ref{fig.1}(d)], under gravity, ${\bm g}$. The system is initialised with a given number of oil droplets, of total area $A_{\text{oil}}$, such that the volume fraction of the initially dispersed phase is $\phi=A_{\text{oil}}/(L H)$.
The intensity of buoyancy forces is encoded in the Rayleigh number
\begin{equation}\label{eq:Ra}
\Ra = \frac{\alpha\, g\, \Delta T\, H^{3}}{\nu\, \kappa},
\end{equation}
where $\nu$, $\kappa$ and $\alpha$ are, respectively, the kinematic viscosity, the thermal diffusivity and the thermal expansion coefficient of 
the continuous phase.
Hereafter, the two fluids are assumed to have identical physical properties (densities, viscosities, thermal diffusivities, and thermal expansion coefficients). Otherwise, the definition of $\Ra$ should incorporate weighted averages based on the mass fractions of the two components. Although emulsions can in principle be formulated with nearly matched thermal expansion coefficients (e.g., paraffin oil in water–ethanol mixtures), the two phases generally exhibit significant differences in $\alpha$. This mismatch affects convective dynamics, as dispersed and continuous phases experience buoyancy forces of different magnitudes, effectively imparting a form of dynamical inertia to the droplets. Owing to its practical relevance, this aspect deserves further dedicated investigation.
As remarked earlier, emulsions display, in general, an effective viscosity that depends on $\phi$ and, for large values of $\phi$, on the shear-rate [see Fig.~\ref{fig.1}(b)]. Hence, $\Ra$ cannot be uniquely defined. It is therefore taken the kinematic viscosity $\nu$ of the continuous phase, and we use $\Ra$, given in Eq.~\eqref{eq:Ra}, as a dimensionless measure of the imposed forcing. Despite the increased complexity of the global energy balance in emulsions, where viscous dissipation is supplemented by interfacial contributions, we characterise heat transport through the dimensionless time-averaged Nusselt number~\cite{Ahlers09} 
\begin{equation}\label{eq:Nu}
\overline{\Nu} = \langle \Nu(t) \rangle_t = \Big\langle 1 + \frac{\langle u_z({\bm x},t)\,T({\bm x},t)\rangle_{x,z}}{\kappa\,(\Delta T/H)}\Big\rangle_t,
\end{equation}
which depends on the hydrodynamical [${\bm u}(x,z,t)$] and temperature [$T(x,z,t)$] fields. The symbol $\langle \dots \rangle_{x,z}$ denotes the spatial average over the whole RB cell and $\langle \dots \rangle_{t}$ the average over the statistically steady state. Indeed, after a transient state, emulsions reach a statistically steady state where the Nusselt number $\Nu(t)$ exhibits fluctuations in time around the mean value $\overline{\Nu}$ [see Fig.~\ref{fig.2}(a)]. To analyse the non-trivial phenomenology emerging at smaller scales, the Nusselt number at the droplet scale is introduced~\cite{PelusiSM21,PelusiSM23,PelusiPRL24,PelusiPRF25}
\begin{equation}\label{eq:NuDrop}
\mathrm{Nu}_{\mathrm{drop}}(t) = 1+\frac{u_{\mathrm{drop},z}(t)  T_{\mathrm{drop}}(t)}{\kappa \frac{\Delta T}{H}},
\end{equation}
where $u_{\mathrm{drop},z}$ and $T_{\mathrm{drop}}$ refer to the vertical velocity and temperature of a given droplet, respectively. Furthermore, we analyse the droplet displacement, {\bf d}$(x,z,t)$ [see Fig.~\ref{fig.1}(d)]. In particular, the x–averaged fluctuations of the displacement with respect to its time-average~\cite{PelusiSM21}
\begin{equation}\label{eq:displ_fluct}
    \tilde{\delta \mathrm{\bf d}}(z,t) = \langle \delta \mathrm{\bf d}(x,z,t)\rangle_x = \big\langle \mathrm{\bf d}(x,z,t) - \langle \mathrm{\bf d}(x,z,t) \rangle_t \big\rangle_x,
\end{equation}
is used to highlight spatio-temporal correlations.
\section{Impact of interfacial stabilisation} 
A fundamental aspect in the study of thermal convection is the characterisation of the heat transferred through the system, quantified by means of the Nusselt number [see Eq.~\eqref{eq:Nu}], in response to an applied buoyancy forcing, namely $\Ra$ [see Eq.~\eqref{eq:Ra}]~\cite{Ahlers09,Chilla12}. In the case of emulsions, the dynamical response also depends crucially on the volume fraction of the initially dispersed phase, $\phi$, which determines how much interface is present and, therefore, affects the energetic balances and the rheology of the material~\cite{PelusiPRE25}.\\
In Fig.~\ref{fig.1}(a)-(b), a set of typical oil-in-water emulsion morphologies obtained from numerical simulations for different values of $\phi$ is reported with the associated rheological flow-curves. 
Under dilute to semi-dilute conditions ($\phi \lesssim 0.5$), the emulsion behaves as a Newtonian fluid with a $\phi$-dependent effective viscosity~\cite{Zinchenko84,Pal2000,PelusiSM21}. For $0.5 \lesssim \phi \lesssim 0.7$, non-Newtonian effects progressively emerge, culminating at high concentrations ($\phi \gtrsim 0.7$) in the appearance of a finite yield stress $\Sigma_Y$. In this regime, the flow curves are well described by the Herschel–Bulkley law, $\Sigma = \Sigma_Y + A\dot{\gamma}^n$ with fit parameters $\Sigma_Y$, consistency index, $A$, and flow index, $n$, that, upon proper normalization, agree with experimental values~\cite{Mason96,Goyon08,PelusiSM21,PelusiPRE25}.
It is important to remark that this complex rheological scenario emerges only if droplets are stabilised against coalescence~\cite{Ravera2021}. In the absence of stabilisation mechanism, a liquid-liquid dispersion with a volume fraction $\phi>0.5$ (i.e., where the majority phase is oil) cannot be supported: the system would always (irrespective of the forcing) undergo an inversion, whereby the initially continuous minority phase (water) becomes the dispersed phase, at a volume fraction 
$1-\phi$ [see Fig.~\ref{fig.1}(c), right panels]. Even for $\phi<0.5$, though, the absence of the stabilisation mechanism has a great impact on the emulsion morphology: in this case, in fact, coalescence events become statistically relevant, and the resulting dynamical equilibrium between breakup and coalescence events gives rise to domains of sizes spanning a broad range of scales. In contrast, in stabilised liquid-liquid dispersions, i.e., in proper emulsions, the droplet size distribution remains slightly polydisperse but peaked at relatively small sizes [see Fig.~\ref{fig.1}(c), left panels].
Despite these structural differences, which are markedly evident in Fig.~\ref{fig.1}(c), the global heat-transfer response remains similar regardless of the nature of the system (i.e., stabilised vs. non-stabilised liquid-liquid dispersion) in dilute conditions~\cite{PelusiPRF25}. As highlighted in Fig.~\ref{fig.2}(b), for small values of $\phi$ (i.e., $\phi=0.16$), the time-averaged Nusselt number 
$\overline{\Nu}$ (see Eq.~\eqref{eq:Nu}) shows a similar behaviour as a function of $\Ra$, for both stabilised (blue circles, top panel) and non-stabilised (yellow circles, bottom panel) systems. At volume fractions $\phi>0.5$ 
(i.e., $\phi=0.84$) the picture changes dramatically. Non-stabilised liquid-liquid dispersions always run into the phase-inverted water-in-oil configuration, and therefore behave as the “complementary” oil-in-water emulsion with a volume fraction $1-\phi$ at any $\Ra$. 
Conversely, stabilised liquid–liquid dispersions undergo a sharp transition from a conductive ($\Nu = 1$) to a convective ($\Nu > 1$) state as $\Ra$ increases, marking the onset of phase inversion at sufficiently strong forcing. This mechanism is further discussed in the section on intermittency and illustrated in Fig.~\ref{fig.4}.

\section{Role of finite-sized droplets}

\begin{figure}
\centering
\includegraphics[width=.95\linewidth]{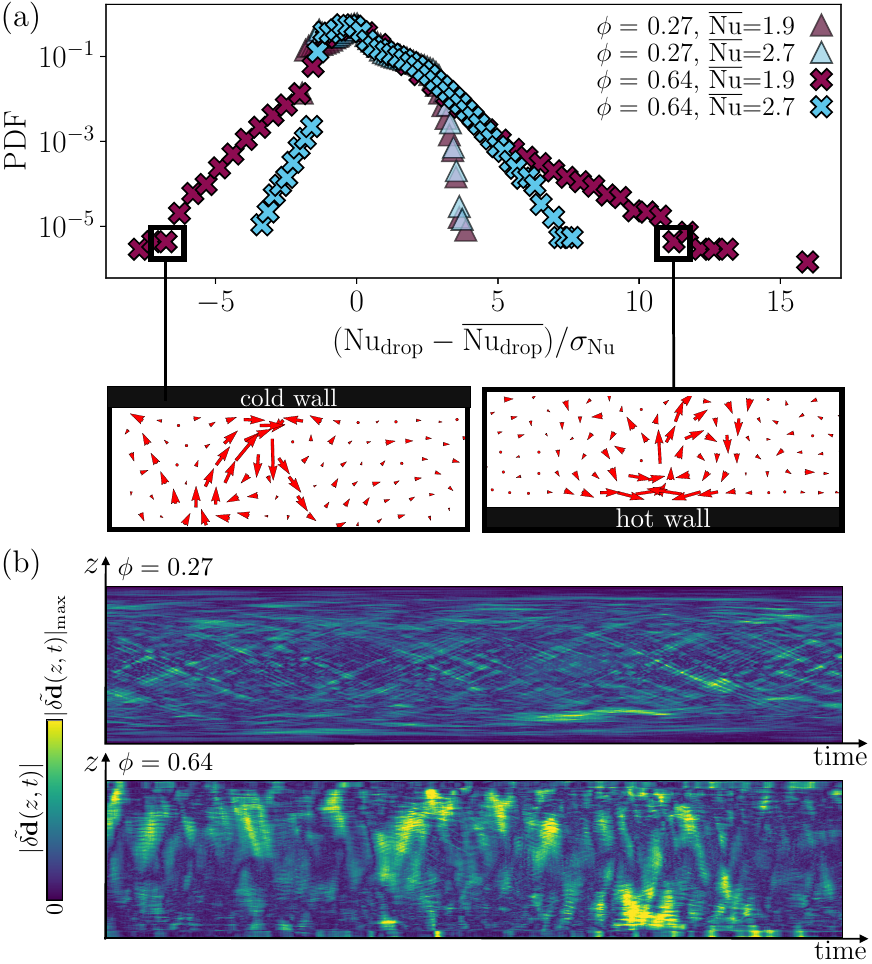}
\caption{Analysis of heat transfer at the droplet scale. Panel (a): Log-lin plot of the PDF of the droplet Nusselt number $\Nu_{\mathrm{drop}}$ [see Eq.~\eqref{eq:NuDrop}], computed over all droplets and time frames. Values are expressed in units of the standard deviation, $\sigma_{\Nu}$, relatively to the mean $\overline{\Nu_{\mathrm{drop}}}$. Different symbols (colors) refer to different values of $\phi$ ($\overline{\Nu}$). We also report zoomed snapshots of displacement fluctuations $\delta \mathrm{\bf d}(x,z,t)$ [see Eq.~\eqref{eq:displ_fluct}] of droplets contributing to positive and negative PDF tails. Panel (b): Spatio–temporal map of the absolute value of the $x$–averaged displacement fluctuations, $\tilde{\delta \mathrm{\bf d}}(z,t)$ [see Eq.~\eqref{eq:displ_fluct}].}
\label{fig.3}
\end{figure}

\begin{figure*}
\centering
\includegraphics[width=.95\linewidth]{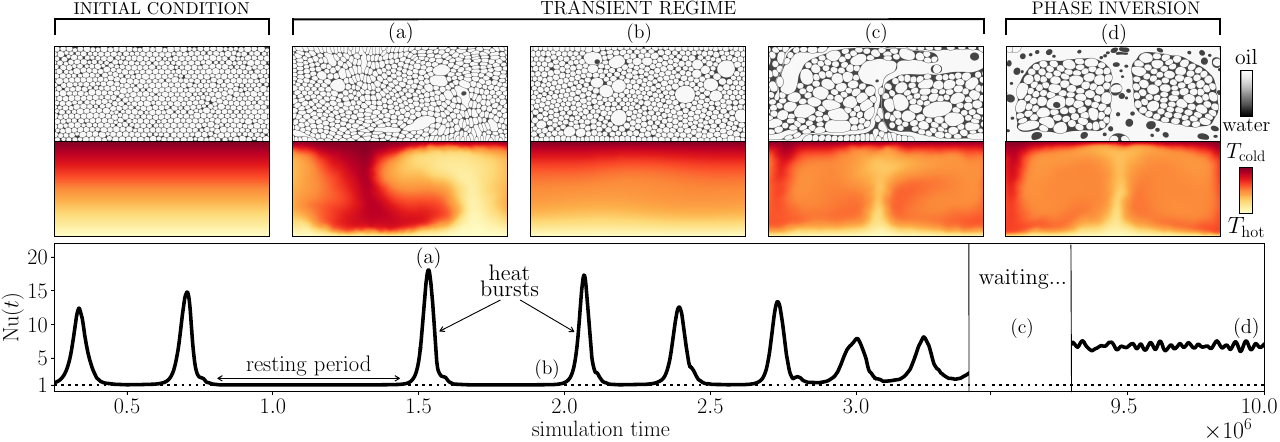}
\caption{Time evolution of the average heat transfer $\Nu(t)$ of a jammed emulsion with $\phi = 0.79$ and $\Ra \approx 4 \times 10^5$. Dotted black line marks the conductive regime ($\Nu = 1$), while upper panels show maps of density (first row) and temperature (second row) for some selected  configurations [panels (a)–(d)]. Time is reported in simulation units.}
\label{fig.4}
\end{figure*}

So far, we have highlighted that, for large values of $\phi$, the suppression of droplet coalescence in emulsions markedly changes the system's heat transfer response compared to non-stabilised liquid-liquid dispersions. In fact, an increase in the number of droplets leads to more frequent collisions, so droplets repeatedly interact without merging, giving rise to localized, small-scale fluctuations in the heat flux. These fluctuations have been observed to be more pronounced as $\phi$ increases and when approaching the transition from conduction to convection from above~\cite{PelusiSM21}. This evidence opened some questions concerning the role played by the presence of finite-sized droplets on the heat transfer properties of emulsions, since heat-flux fluctuations at the droplet scale respond strongly to the stabilisation mechanism: they are amplified due to sustained droplet interactions, while coalescence suppresses them in non-stabilised systems for $\phi \geq 0.2$~\cite{PelusiPRF25}. To investigate this aspect, it is necessary to scale down the analysis to the droplet level. 
For this analysis, the TLBfind code~\cite{TLBfind22} proved to be an optimal tool of investigation, since it is equipped with a Lagrangian analysis tool that allows identifying droplets, their centres of mass, and following their trajectories over time. We collected the values of the droplet-scale Nusselt number $\Nu_{\mathrm{drop}}$ [see Eq.~\eqref{eq:NuDrop}] for all droplets at all times during the statistically steady regime, and we estimated the probability density functions (PDF) of the fluctuations of $\Nu_{\mathrm{drop}}$ around the mean value, for two volume fractions ($\phi=0.27$ and $\phi=0.64$) at fixed intensity of the global heat transfer [i.e., fixed $\overline{\Nu}$, see Fig.~\ref{fig.3}(a)]. PDFs exhibit markedly broader non-Gaussian tails, which become more pronounced as $\phi$ increases. These tails indicate bursts of heat transport, associated to local fluidisation of the material, which arises from the coupling between small-scale velocity fluctuations and droplet–droplet collision dynamics, enabled by the suppression of coalescence. Notice that pronounced droplet-scale heat-flux fluctuations appear only in the presence of finite-size droplets and are caused by droplets moving toward or away from the walls [see snapshots in Fig.~\ref{fig.3}(a)]~\cite{PelusiSM21,PelusiSM23}. In fact, as $\phi$ increases, spatial correlations among droplets become stronger; in contrast, in dilute emulsions, droplets' correlations weaken since perturbations generated by droplets' displacement cannot propagate throughout the system. This aspect is confirmed by Fig.~\ref{fig.3}(b), reporting the spatio-temporal dynamics of the $x$-averaged droplet displacement fluctuations $\tilde{\delta \mathrm{\bf d}}(z,t)$ [see Eq.~\eqref{eq:displ_fluct}]: only if the system is sufficiently packed, then displacement fluctuations depart from zero coherently in extended space regions located close to the boundaries. On the contrary, space-time coherence is visibly lost when $\phi$ lays in the dilute regime.
\section{Intermittency and phase inversion in jammed emulsions}
As mentioned earlier, if $\phi$ is increased, the emulsion rheology develops a yield-stress [see
Fig.~\ref{fig.1}(b)], which strongly hinders convection~\cite{Darbouli13,Turanetal12}.
At sufficiently large values of $\Ra$, the system turns into a convection regime, but with distinctively peculiar phenomenology, in comparison to what is observed in the absence of yield stress. In Fig.~\ref{fig.4}, we report the Nusselt number $\Nu(t)$ for an emulsion with $\phi=0.79$: the transient dynamics of the heat transfer strongly differs from that of dilute cases [see Fig.~\ref{fig.2}(a)], since it is characterised by intense convective bursts (whereby $\Nu \gg 1$), occurring intermittently in an otherwise conductive state ($\Nu \approx 1$). During the conductive (``resting") periods, the emulsion is not completely quiescent since rearrangements of the local droplet topology occur, relaxing the stored elastic energy. It is, indeed, this residual plastic activity which triggers further ``heat bursts". It is important to remark that this mechanism, which is crucial to sustain the intermittent convection, is tightly related to the inherent ``granularity" of the emulsion structure. The bursting events are associated with a transient fluidisation of the material, with the emergence of the typical convective-roll structure of the velocity field, whose strong gradients promote droplet coalescence (see snapshots in Fig.~\ref{fig.4}). It is worth noting that continuum models with local constitutive relations between stress and shear rate hardly capture the above mentioned intermittent behaviour~\cite{Zhang06,BalmforthRust09,Turanetal12}.
Once convection stops during the ``resting periods", these models offer no mechanism to spontaneously restart it, since they lack the key ingredient, that is the discrete nature of the emulsion, i.e., finite-sized droplets undergoing plastic rearrangements~\cite{Goyon08}.
Moreover, convective periods modify the emulsion morphology by promoting coalescence, which reduces the total interfacial area. As a result, the system becomes less robust to stress and more susceptible to fluidisation and subsequent heat bursts, whose frequency increases over time.
Eventually, this process leads to a spatially heterogeneous state, whereby islands of the original concentrated (yield-stress) emulsion are embedded in a matrix of its dilute (Newtonian) phase-inverted counterpart [see snapshots in Fig.~\ref{fig.4}]. Moreover, once this partial phase inversion has taken place, the heat dynamics enters a sustained convection state, typical of Newtonian fluids at moderate Rayleigh numbers, as one can appreciate from the fluctuating, statistically steady, signal of $\overline{\Nu}$ in the rightmost part of the graph of Fig.~\ref{fig.4}. Notice that phase inversion is only one of the possible dynamical regimes which emulsions explore for different combinations of the pair ($\phi$, $\Ra$), since also some breakup- or coalescence-dominated regimes can be reached~\cite{PelusiPRE25}. We remark that the morphological transition from an oil-in-water to a water-in-oil emulsion induced by phase inversion is irreversible~\cite{Bouchama03,Yi24}. Indeed, as previously highlighted when discussing the $\overline{\Nu}$ vs. $\Ra$ curve for highly concentrated emulsions, once the phase inversion occurs, the system cannot recover its original oil-in-water configuration upon decreasing the buoyancy forcing. Consequently, the $\overline{\Nu}$ vs.$\Ra$ curve no longer exhibits the jump (hysteresis).
\section{Conclusions and Perspectives}
Emulsions are soft materials consisting of the dispersion of finite-sized droplets, whose complex rheology depends on the dispersed-phase volume fraction, ranging from Newtonian (dilute) to non-Newtonian (concentrated) behaviour.
The coupling between this structural/rheological complexity and convective dynamics opens up a fascinating and previously unexplored phenomenology, particularly for: \textit{i)} large values of $\phi$, resulting in non-Newtonian emulsions~\cite{Bonn17}, and \textit{ii)} interfacial dynamics involving breakup and coalescence, favouring structural variations~\cite{Stone94,Perlekar2012,Girotto22,Girotto24,Roccon17,Brandt24,Mangani24,PelusiPRE25}. In this perspective, we have highlighted distinctive hallmarks of this novel phenomenology, featuring abrupt transitions in the $\overline{\Nu}$ vs.~$\Ra$ relation associated with mesoscale plasticity~\cite{PelusiPRE25}, morphology evolution towards phase-inverted states induced by dynamics, and long-lasting transient states with intermittent heat transfer properties~\cite{PelusiPRL24,PelusiPRE25}.\\
Numerical simulations are essential to reveal the emulsion behaviour across different spatial and temporal scales; in particular, the methodology employed here~\cite{TLBfind22} is well suited to simulate emulsion fluid dynamics under RB thermal convection, whereas experiments face, in fact, intrinsic difficulties, such as tracking the breakup-coalescence dynamics and droplet-size distributions (due to the optical opacity of dense dispersions which often hinders direct flow visualization~\cite{Hu17}), or 
achieving precise density matching to prevent buoyancy-driven creaming~\cite{chaochat}.  
Indeed, controlled experiments on RB convection in jammed emulsions are, to the best of our knowledge, missing in the literature. Numerical simulations could, then, provide guidelines for their design and pointers to the regions of parameter space where novel and interesting effects can be detected. For example, observing the intermittency phenomenon reviewed in Fig.~\ref{fig.4} would require an emulsion with relatively small value of the yield stress, with respect to the buoyancy strength (i.e., a small Bingham number). A concentrated oil-in-water emulsion, with, e.g., silicone oil as the oil phase, can develop a yield stress as low as $\Sigma_Y \approx 0.1 \, \text{Pa}$ (as obtained in Ref.~\cite{Mason96} for $\phi \approx 0.65$), such that an experimentally realistic RB cell of width $H \approx 0.5 \, \text{m}$ with a temperature difference of $\Delta T \approx 2 \, \text{K}$ (which also allows to avoid unwanted strong thermocapillary effects), would be suitable for the scope.
In this perspective, it is important to note that all numerical computations discussed have been carried out in 2D due to the need for an appropriate resolution to capture droplet dynamics and obtain a reasonable amount of statistics. While 2D convection can retain essential features of 3D convection, at least concerning global large-scale quantities~\cite{van2013comparison}, the computational challenges posed by 3D numerical simulations are, then, worth being embraced and are actually a matter of presently ongoing work. 
\acknowledgments
Prof. Chao Sun is kindly acknowledged for useful discussions. FP acknowledges the Innovative Medicines Initiative 2 Joint Undertaking (JU), grant agreement No 853989. The JU receives support from the European Union’s Horizon 2020 Research and Innovation Programme and EFPIA and Global Alliance for TB Drug Development Non-Profit Organisation, Bill \& Melinda Gates Foundation, University of Dundee. MS and MB acknowledge the support of the National Center for HPC, Big Data and Quantum Computing, Project CN\_00000013 – CUP E83C22003230001 and CUP B93C22000620006, Mission 4 Component 2 Investment 1.4, funded by the European Union – NextGenerationEU. Support from INFN/FIELDTURB project is also acknowledged.

\subsection{Data availability statement}
The data that support the findings of this study are available upon reasonable request from the authors.

\bibliography{francesca.bib}

\end{document}